\documentclass[pra,twocolumn,showpacs,preprintnumbers,amsmath,amssymb]{revtex4}
\usepackage{mathrsfs}
\usepackage{bbm}
\usepackage{amsfonts}
\usepackage{tipa}

\usepackage{epsfig,graphicx}
\usepackage{amstext}
\usepackage{amsmath}
\usepackage{graphicx}

\begin{document}


\title{Quantum-memory-assisted entropic uncertainty principle, teleportation and
       entanglement witness in structured reservoirs}

\author{Ming-Liang Hu$^{1,}$}
\email{mingliang0301@163.com}
\author{Heng Fan$^{2,}$}
\email{hfan@iphy.ac.cn}
\address{$^{1}$School of Science, Xi'an University of Posts and
               Telecommunications, Xi'an 710121, China \\
         $^{2}$Beijing National Laboratory for Condensed Matter Physics,
               Institute of Physics, Chinese Academy of Sciences, Beijing
               100190, China}

\begin{abstract}
We relate the principle of quantum-memory-assisted entropic uncertainty
to quantum teleportation and show geometrically that any two-qubit
state which lowers the upper bound of this uncertainty relation is
useful for teleportation. We also explore the efficiency of this
entropic uncertainty principle on witnessing entanglement in a
general class of bosonic structured reservoirs. The
entanglement regions witnessed by different estimates are determined,
which may have no relation with the explicit form of the spectral density of
the reservoir for certain special chosen sets of the initial states.
\end{abstract}


\maketitle

\section{Introduction}
The uncertainty principle initially observed by Heisenberg
\cite{Heisenberg} for the case of position and momentum, and further
formulated by Robertson \cite{Robertson} for arbitrary pairs of
observables is a central in quantum theory. It sets
limits on our ability to predict the precise outcomes of two
incompatible measurements on a quantum system, and at the same time
provides the basis for new technologies such as quantum cryptography
\cite{Bennett} in quantum information. Originally, the uncertainty relation is expressed in
terms of the standard deviation $\Delta R\cdot\Delta
S\geqslant\frac{1}{2}|\langle[R,S]\rangle|$ for two observables $R$
and $S$ \cite{Robertson}. However, this uncertainty bound is
state-dependent and also trivial for finite-valued observables
\cite{Prevedel}. To remove this pitfall and to precisely capture its
physical meanings, the original form has subsequently been recast to
the entropic one \cite{Deutsch}, which reads
\begin{equation}
 H(R)+H(S)\geqslant \log_2 \frac{1}{c},
\end{equation}
where $H(R)$ denotes the Shannon entropy of the probability
distribution of the outcomes when $R$ is measured, and likewise for
$H(S)$. $1/c$ quantifies the complementarity of $R$ and $S$, where
$c=\max_{r,s}|\langle \Psi_r|\Phi_s\rangle|^2$ for nondegenerate
observables, with $|\Psi_r\rangle$ and $|\Phi_s\rangle$ being the
eigenvectors of $R$ and $S$.

The above uncertainty limitation applies to the case that the
observer can access only to the classical information. It may be
violated through clever use of entanglement, which plays a central
role in quantum information. This is the quantum-memory-assisted
entropic uncertainty principle, which was previously conjectured by
Renes and Boileau \cite{Renes}, and later being strictly proved by
Berta {\it et al.} \cite{Berta}. It states that if the observer can
entangle the particle $A$ that he wishes to measure with another
particle $B$ which serves as a quantum memory, then the uncertainty
of this observer about any pair of observables can be dramatically
reduced. Particularly, if $A$ and $B$ are maximally entangled, then
the observer is able to correctly predict the outcomes of whichever
measurement is chosen. This new entropic uncertainty relation reads
\begin{equation}
 H(R|B)+H(S|B)\geqslant \log_2 \frac{1}{c}+H(A|B),
\end{equation}
where $H(R|B)$ is the conditional von Neumann entropy of the
postmeasurement state
$\rho_{RB}=\sum_{r}(|\Psi_r\rangle\langle\Psi_r|\otimes
\mathbb{I})\rho_{AB}(|\Psi_r\rangle\langle\Psi_r|\otimes
\mathbb{I})$, and likewise for $H(S|B)$. Compared with Eq. (1),
there is an extra term $H(A|B)$ (the conditional von Neumann entropy
of $\rho_{AB}$) appearing on the right-hand side of Eq. (2). In
particular, if $H(A|B)$ is negative, the Berta {\it et al.} bound
(BB) of uncertainty in Eq. (2) can be reduced comparing with
previous uncertainty relation. It is also found that this negative
value gives the lower bound of the one-way distillable entanglement
between $A$ and $B$ \cite{Devetak}.

This new entropic uncertainty principle has been recently confirmed
experimentally \cite{Prevedel,Licf}, and ignites interests of people
on investigating its potential applications from various aspects
\cite{Coles}. In this work, we will first relate it to quantum
teleportation, and show that any two-qubit $\rho_{AB}$ with negative
$H(A|B)$ gives nonclassical teleportation fidelity. It is known that
teleportation is one fundamental protocol in quantum information
processing \cite{Bennettprl}. It is crucial, both theoretically and
experimentally, to know whether the fidelity of teleportation is in
classical regime or in quantum regime. Our result relate this
important problem to the entropic uncertainty principle.

We will also investigate efficiency of this new entropic uncertainty
relation on witnessing entanglement in a class of bosonic structured
reservoirs. It is known that entanglement plays a key role in
quantum information processing such as in teleportation and in
condensed matter physics; see, for example, Refs.
\cite{nature1,naturec}. The point of departure for this practical
application (i.e., entanglement witness) is Eq. (2), from which one
can note that if $H(R|B)+H(S|B)<\log_2(1/c)$, then $H(A|B)<0$, and
hence $\rho_{AB}$ is entangled \cite{Devetak}. Experimentally, the
value of $H(R|B)+H(S|B)$ can be estimated by conditional
single-qubit tomography on $B$ \cite{Prevedel}, and this estimate is
termed tomographic estimate (TE).

There are other ways for estimating the uncertainty. The first is
the measurement estimate (ME) denoted by $H(R|R)+H(S|S)$, which
corresponds to the same measurements on both $A$ and $B$, and are
favored for its ease of implementation \cite{Prevedel}. This
estimate provides an upper bound for the new uncertainty relation in
that quantum measurements never decrease entropy. The second is the
Fano estimate (FE) obtained by using Fano's inequality
$H(X|B)\leqslant h(p_X)+p_X\log_2(d-1)$ \cite{Nielsen}, where
$h(p_X)$ is the binary entropy function, with $p_X$ being the
probability that the outcomes of $X$ on $A$ and $X$ on $B$ are
different, and $d$ is the dimension of $A$. For the two-qubit system
(i.e., $d=2$), the inequality $h(p_R)+h(p_S)<\log_2(1/c)$ is a
signature of entanglement between $A$ and $B$.

\section{Linking the new entropic uncertainty relation to teleportation}
In this section we relate the quantum-memory-assisted entropic
uncertainty principle to quantum teleportation. We will show that
any two-qubit state with negative conditional von Neumann entropy,
which thus lowers the upper bound of the uncertainty, is a
manifestation of its usefulness for nonclassical teleportation.

Without loss of generality, we suppose the sender Alice wants to
teleport to the receiver Bob a general one-qubit state, with a
two-qubit state $\tau_{AB}$ (pure or mixed) being used as the
quantum channel. Then if they adopt the standard teleportation
scheme (i.e., Alice performs the Bell-basis measurement while Bob is
equipped to perform any unitary transformation), the maximal average
fidelity achievable can be evaluated as \cite{Horodeckiletta}
\begin{equation}
 F_{\rm av}=\frac{1}{2}+\frac{1}{6}N(\tau_{AB}),
\end{equation}
where $N(\tau_{AB})={\rm tr}\sqrt{T^\dag T}$, with $T$ being the
$3\times 3$ positive matrix with elements $t_{ij}$ related to the
Bloch sphere representation of $\tau_{AB}$ below
\begin{equation}
 \tau_{AB}=\frac{1}{4}(\mathbb{I}\otimes\mathbb{I}+\vec{x}\cdot\vec{\sigma}\otimes\mathbb{I}+
 \mathbb{I}\otimes\vec{y}\cdot\vec{\sigma}+\sum_{i,j=1}^3 t_{ij}\sigma_i\otimes\sigma_j),
\end{equation}
where $\mathbb{I}$ is the $2\times2$ identity operator,
$\vec{\sigma}=(\sigma_1,\sigma_2,\sigma_3)$ is the vector of the
Pauli spin matrices, $\vec{x}=(x_1,x_2,x_3)$ and
$\vec{y}=(y_1,y_2,y_3)$ are the local Bloch vectors in
$\mathbb{R}^3$ with $\vec{x}\cdot\vec{\sigma}=\sum_{i=1}^{3}x_i
\sigma_i$ and $\vec{y}\cdot\vec{\sigma}=\sum_{i=1}^{3}y_i \sigma_i$.

Since the teleportation fidelity is locally unitary invariant, and
we can always find unitary operators $\mathcal {U}_A$ and $\mathcal
{U}_B$ which transform the state $\tau_{AB}$ into
\begin{equation}
\rho_{AB}=\frac{1}{4}(\mathbb{I}\otimes\mathbb{I}+\vec{r}\cdot\vec{\sigma}\otimes\mathbb{I}+
 \mathbb{I}\otimes\vec{s}\cdot\vec{\sigma}+\sum_{k=1}^3 v_{k}\sigma_k\otimes\sigma_k),
\end{equation}
with $\vec{r}=(r_1,r_2,r_3)$ and $\vec{s}=(s_1,s_2,s_3)$, it
suffices to restrict our concern to the representative class of
quantum channels expressed in Eq. (5) with less number of
parameters, for which we always have
\begin{equation}
 N(\rho_{AB})=\sum_{k=1}^{3} |v_k|.
\end{equation}
$N(\rho_{AB})>1$ gives $F_{\rm av}>2/3$ and thus $\rho_{AB}$ is
competent for teleportation.

\begin{figure}
\centering
\resizebox{0.4\textwidth}{!}{%
\includegraphics{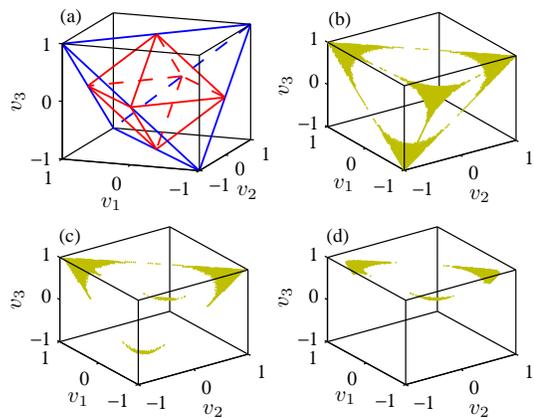}}
\caption{(Color online) Geometry of the tetrahedron $\mathcal {T}$
and the octahedron $\mathcal {O}$ associated with the vector
$\vec{v}$ of $\rho_{AB}$ (a), and valid regions of $\vec{v}$
constrained by $H(A|B)<0$ for $\vec{r}=\vec{s}=(0,0,0)$ (b),
$(0,0,0.25)$ (c), and $(0.1,0.1,0.25)$ (d).} \label{fig:1}
\end{figure}

The real numbers $r_k$, $s_k$, and $v_k$ ($k=1,2,3$) in Eq. (5) must
satisfy certain constraints such that $\rho_{AB}$ is a well defined
density operator. Particularly, for the special case of
$\vec{r}=\vec{s}=0$, physical $\rho_{AB}$ are those with
$(v_1,v_2,v_3)$ (for simplicity, we just refer to $\vec{v}$ in the
following) belongs to the tetrahedron $\mathcal {T}$ (see Fig. 1)
with vertices $(-1,1,1)$, $(1,-1,1)$, $(1,1,-1)$ and $(-1,-1,-1)$,
among which separable ones are confined to the octahedron $\mathcal
{O}$ with vertices $(\pm 1,0,0)$, $(0,\pm 1,0)$, and $(0,0,\pm1)$
\cite{Horodecki}. The Bell states sit at the four vertices of
$\mathcal {T}$, while the Werner states \cite{Ewerner} are those
represented by the lines connecting the vertices of $\mathcal {T}$
with the origin of $\mathcal {O}$.

For the nonzero values of $\vec{r}$ and/or $\vec{s}$, the vector
$\vec{v}$ which admits the positive semidefiniteness of $\rho_{AB}$
also belongs to $\mathcal {T}$ \cite{Horodecki}. But now some
$\vec{v}$ inside $\mathcal {T}$ may not correspond to the physical
states. Moreover, while all the separable states are still confined
to $\mathcal {O}$, there are also entangled ones with $\vec{v}$
belongs to $\mathcal {O}$, which is different from the case of
$\vec{r}=\vec{s}=0$.

Since the octahedron $\mathcal {O}$ is specified by $\sum_{k=1}^{3}
|v_k|\leqslant1$, one can see from Eq. (6) that any physical
$\rho_{AB}$ with $\vec{v}$ belongs to the four small tetrahedra
divided by $\mathcal {O}$ (i.e., the regions inside $\mathcal {T}$
and outside $\mathcal {O}$) gives $N(\rho_{AB})>1$ and thus is
useful for quantum teleportation. Physical $\rho_{AB}$ confined to
$\mathcal {O}$ may also be entangled when the inner product
$(\vec{r},\vec{s})\neq 0$, but they are useless for teleportation.

Now we begin to discuss the operational identification of
$\rho_{AB}$ useful for teleportation, and to what degree this
identification can cover the whole set of physical $\rho_{AB}$
useful for teleportation. When considering the standard protocol,
Horodecki {\it et al.} proved that any $\rho_{AB}$ which violates
the Bell-Clauser-Horne-Shimony-Holt (Bell-CHSH) inequality is useful
for teleportation \cite{Horodeckiletta}. Here, we will show that the
negativity of the conditional von Neumann entropy of $\rho_{AB}$ is
also a signature of its usefulness for teleportation. Particularly,
these $\rho_{AB}$ can be witnessed by means of the new entropic
uncertainty relation, and therefore is expected to have potential
applications in experiments.

To prove the above argument, we consider first $\rho_{AB}$
associated with the vertices of $\mathcal {O}$. The positive
semidefiniteness requires that the matrix elements
$\rho_{AB}^{11}\rho_{AB}^{44}\geqslant |\rho_{AB}^{14}|^2$ and
$\rho_{AB}^{22}\rho_{AB}^{33}\geqslant |\rho_{AB}^{23}|^2$, which
yield $r_{3}=s_{3}=0$ for physical states. For example, if
$\vec{v}=(1,0,0)$ the above requirements turn to
$1-(r_3+s_3)^2\geqslant 1$ and $1-(r_3-s_3)^2\geqslant 1$, which are
satisfied only when $r_{3}=s_{3}=0$.

Under the conditions of $r_{3}=s_{3}=0$, we can write the explicit
forms of $\rho_{AB}$ at the vertices of $\mathcal {O}$, and further
determine the constraints imposed on the parameters by positive
semidefiniteness of $\rho_{AB}$. We use another property of the
physical $\rho_{AB}$, which says that if a Hermitian matrix is
positive semidefinite then all of its principal minors must be
non-negative.

For $\rho_{AB}$ with $\vec{v}=(1,0,0)$, we derive the second- and
third-order leading principal minors as
\begin{eqnarray}
D_2=\frac{1-s_1^2-s_2^2}{16},~
D_3=-\frac{(r_1-s_1)^2+(r_2-s_2)^2}{64}.
\end{eqnarray}
We see that $s_1^2+s_2^2\leqslant 1$, $r_1=s_1$ and $r_2=s_2$ must
be satisfied for ensuring the positive semidefiniteness of
$\rho_{AB}$. Furthermore, the determinant of the $(3,3)$ minor
formed by removing from $\rho_{AB}$ its third row and third column
(i.e., one of the third-order principal minor) can be determined as
$\Delta_3=-r_2^2/16$, $\Delta_3\geqslant 0$ further gives $r_2=0$.
Then under the conditions of $r_{2,3}=s_{2,3}=0$ and $-1\leqslant
r_1=s_1\leqslant 1$ we obtain the eigenvalues of $\rho_{AB}$ as
$\epsilon_{1,2}=0$, $\epsilon_{3,4}=(1\pm r_1)/2$, and the
eigenvalues of the reduced $\rho_B={\rm tr}_A\rho_{AB}$ as
$\varepsilon_{1,2}=(1\pm r_1)/2$. These give rise to the quantum
conditional entropy $H(A|B)=0$.

In fact, one can also determine $r_2=0$ by the argument that
$\rho_{AB}$ is positive semidefinite if ${\rm tr}(\rho_{AB}\mathcal
{P})\geqslant 0$ for any projector $\mathcal {P}$. Taking $\mathcal
{P}=uu^\dag$ with $u=(u_1,u_2,u_3,u_4)^T$ and $T$ denoting
transpose, we obtain
\begin{eqnarray}
{\rm
tr}(\rho_{AB}\mathcal{P})&=&\frac{1}{4}(|u_{14}^+|^2+|u_{23}^+|^2)
                                +\frac{1}{2}[r_1{\rm Re}(u_{14}^{+}u_{23}^{+*})\nonumber\\
                                &&-r_2{\rm Im}(u_{14}^{-}u_{23}^{+*})],
\end{eqnarray}
where $u_{ij}^{\pm}=u_i\pm u_j$, with ${\rm Re}(f)$ and ${\rm
Im}(f)$ representing the real and imaginary parts of $f$. One can
check directly that only when $r_2=0$ can ${\rm
tr}(\rho_{AB}\mathcal {P})\geqslant 0$ for any $\mathcal {P}$.

By using the same methodology we obtain constraints imposed on the
parameters of $\rho_{AB}$ associated with the remaining five
vertices of $\mathcal {O}$, they are: $r_{1,3}=s_{1,3}=0$ and
$-1\leqslant r_2=\pm s_2\leqslant 1$ for $\vec{v}=(0,\pm1,0)$,
$r_{1,2,3}=s_{1,2,3}=0$ for $\vec{v}=(0,0,\pm 1)$,
$r_{2,3}=s_{2,3}=0$ and $-1\leqslant r_1=-s_1\leqslant 1$ for
$\vec{v}=(-1,0,0)$. All of these correspond to physical $\rho_{AB}$
with $H(A|B)=0$.

On the other hand, physical states $\rho_{AB}$ with $\vec{v}$
belonging to $\mathcal {O}$ can always be written as a convex
combination of states with $\vec{v}$ at the vertices of $\mathcal
{O}$, so by using the concavity of the quantum conditional entropy
\cite{Nielsen}, we obtain $H(A|B)\geqslant 0$ for any density matrix
$\rho_{AB}$ belongs to $\mathcal {O}$ [it is also possible for
$H(A|B)\geqslant 0$ with $\rho_{AB}$ lying beyond $\mathcal {O}$].
This means that for any physical $\rho_{AB}$ with negative
conditional entropy, $\vec{v}$ must belongs to the four tetrahedra
separated by $\mathcal {O}$, which gives $\sum_{k=1}^3|v_k|>1$, and
thus makes it useful for nonclassical teleportation.

We would like to point out here that the negativity of the
conditional von Neumann entropy and the violation of the Bell-CHSH
inequality \cite{Horodeckiletta} identify different subsets of
density matrices useful for teleportation; namely, there are
$\rho_{AB}$ with $H(A|B)<0$ but do not violate the Bell-CHSH
inequality [e.g., $\rho_{AB}$ of Eq. (5) with
$\vec{r}=\vec{s}=(0,0,0.25)$ and $\vec{v}=(\pm 0.95,\mp
0.25,0.30)$], while there are also $\rho_{AB}$ which violate the
Bell-CHSH inequality but with $H(A|B)>0$ (e.g., the partial of the
extended Werner-like states \cite{Ewerner}).

In Fig. 1 we presented regions of the valid $\vec{v}$ determined by
$H(A|B)<0$ with different $\vec{r}$ and $\vec{s}$, from which one
can see that for the Bell-diagonal states (i.e.,
$\vec{r}=\vec{s}=0$), they locate near the four vertices of
$\mathcal {T}$ and are symmetric with respect to the origin of
$\mathcal {O}$. For the general case $(\vec{r},\vec{s})\neq 0$,
however, the valid $\vec{v}$ makes $H(A|B)<0$ will shrink to small
regions and their distribution are not symmetric with respect to the
origin of $\mathcal {O}$.

The geometric characterization of $\rho_{AB}$ also allows us to
determine fractions of different kinds of $\rho_{AB}$ over the
ensemble of physical $\rho_{AB}$. This can be estimated by
calculating ratio of volumes of the three-dimensional spaces
occupied by $\vec{v}$ associated with different $\rho_{AB}$. Here we
obtained the corresponding volumes by performing Monte Carlo
simulations, generating $10^9$ random $\vec{v}$ uniformly
distributed in the cube illustrated in Fig. 1(a), and checking if
they correspond to physical $\rho_{AB}$, if they give rise to
$F_{\rm av}>2/3$ and if they make $H(A|B)<0$. In this way we
confirmed that $50\%$ of the Bell-diagonal states $\rho_{AB}$ give
$F_{\rm av}>2/3$ (the volume of $\mathcal {T}$ is twice that of
$\mathcal {O}$), and about $4.17\%$ of the Bell-diagonal $\rho_{AB}$
make $H(A|B)<0$. Therefore, about $8.34\%$ of the Bell-diagonal
$\rho_{AB}$ useful for teleportation can be identified by negativity
of the conditional entropy. Using the same method, we have also
performed simulation for $\vec{r}$ and $\vec{s}$ chosen in Fig. 1(c)
and 1(d), and confirmed that for the former (latter) case about
$41.68\%$ ($35.51\%$) of the physical $\rho_{AB}$ are useful for
teleportation, among which about $6.16\%$ ($3.30\%$) of them have
negative quantum conditional entropy.

\section{Entanglement witness in structured reservoirs}
Entanglement is a precious resource for quantum computing, but it is
fragile and can be easily destroyed by the environment. The measure
of entanglement including the operational methods to distinguish it
from the separable case, however, is a tricky problem. So it is of
practical significance to find a straightforward witnessing method.
Here, we discuss efficiency of the new entropic uncertainty relation
on witnessing entanglement in open quantum systems. We consider a
system consists of two identical qubits which interact independently
with their own reservoir, with the single ``qubit+reservoir''
Hamiltonian given by \cite{Breuer}
\begin{equation}
 H=\omega_0\sigma_{+}\sigma_{-}
   +\sum_{k}\omega_k b_k^{\dag} b_k+\sum_{k}(g_k b_k\sigma_{+}+{\rm h.c.}),
\end{equation}
where $\omega_0$ is the transition frequency of the qubit, and
$\sigma_\pm$ are the Pauli raising and lowering operators. The index
$k$ labels the reservoir field mode with frequency $\omega_k$, with
$b_k^{\dag}$ ($b_k$) being the bosonic creation (annihilation)
operator and $g_k$ being the coupling strength.

When the reservoir is at zero temperature and there is no
correlation between the qubit and the reservoir initially, the
single-qubit reduced density matrix $\rho^S(t)$ can then be
determined as \cite{Breuer}
\begin{equation}
 \rho^S(t)=\left(\begin{array}{cc}
  \rho^S_{11}(0)|p(t)|^2  & \rho^S_{10}(0)p(t) \\
  \\
  \rho^S_{01}(0)p^*(t) & 1-\rho^S_{11}(0)|p(t)|^2
 \end{array}\right),
\end{equation}
where $\rho^S_{ij}(0)=\langle i|\rho^S(0)|j\rangle$ in the standard
basis $\{|1\rangle,|0\rangle\}$, and the explicit time dependence of
the single function $p(t)$ contains the information on the reservoir
spectral density and the coupling constants.

After obtaining $\rho^S(t)$, the two-qubit density matrix $\rho(t)$
can then be determined by the procedure presented in
\cite{Bellomoprl}. Here we suppose the two qubits are prepared
initially in the extended Werner-like (EWL) states \cite{Ewerner}
\begin{eqnarray}
 \rho^{\Xi}(0)=r|\Xi\rangle\langle\Xi|+\frac{1-r}{4}\mathbb{I},
\end{eqnarray}
where $|\Xi\rangle=|\psi\rangle$ or $|\phi\rangle$, with
$|\psi\rangle=\alpha|00\rangle+e^{i\theta}\sqrt{1-\alpha^2}|11\rangle$
and
$|\phi\rangle=\alpha|10\rangle+e^{i\theta}\sqrt{1-\alpha^2}|01\rangle$.

The density matrix $\rho^{\Xi}(t)$ depends only on the chosen
initial state $\rho^{\Xi}(0)$ and values of the function $p(t)$
associated with the Hamiltonian model of Eq. (9), regardless of the
reservoir structure. Thus in the following we ignore temporarily the
explicit form of $p(t)$ and consider only the dependence of the
uncertainties on $p$. The results obtained here thus apply to all
cases where the single-qubit dynamics has the form of Eq. (10).

\begin{figure}
\centering
\resizebox{0.4\textwidth}{!}{%
\includegraphics{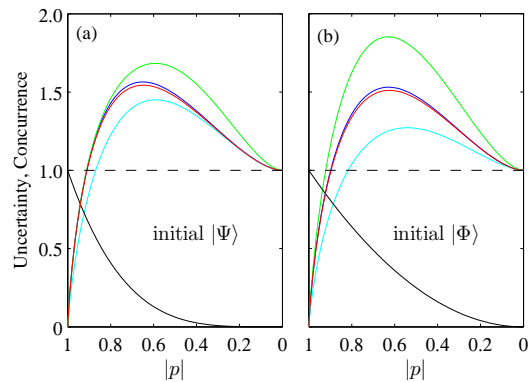}}
\caption{(Color online) The $|p|$ dependence of FE, ME, TE, BB (from
top to bottom) and concurrence (the bottommost) for the initial
$|\Psi\rangle$ and $|\Phi\rangle$. The lines for FE and ME in panel
(a) are plotted by choosing $p\in\mathbb{\mathbb{R}}$.}
\label{fig:2}
\end{figure}

\begin{figure}
\centering
\resizebox{0.4\textwidth}{!}{%
\includegraphics{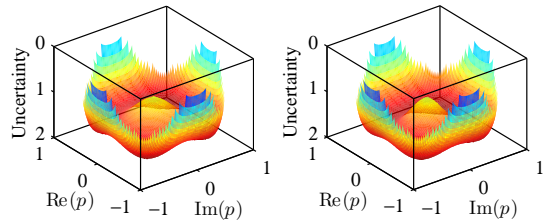}}
\caption{(Color online) Dependence of the ME (left) and FE (right)
on the real and imaginary parts of $p$ for the initial
$|\Psi\rangle$.} \label{fig:3}
\end{figure}

To witness the smallest amount of entanglement
\cite{Prevedel,Berta}, we choose the two observables as $R=\sigma_1$
and $S=\sigma_3$ in the following discussion. This choice of
measurement operators gives the maximal complementarity between $R$
and $S$; that is, $c=1/2$ and thus $\log_2(1/c)=1$.

Consider first the initial state $\rho^{\psi}(0)$, for which the
nonzero elements of $\rho^{\psi}(t)$ can be expressed in terms of
$|p|^2$ (the diagonal elements) or $p^2$ (the antidiagonal elements)
\cite{Bellomoprl,Bellomopra}. Then one can check directly that
uncertainty of the TE, BB as well as the concurrence (a measure of
entanglement) \cite{Wootters} are determined only by $|p|^2$ (they
are also independent of $\theta$), while the other two estimates
(i.e., ME and FE) are determined by $p^2$.

In Fig. 2(a) we plot the uncertainties versus $|p|$ (the ME and FE
are plotted only with $p\in\mathbb{R}$), while in Fig. 3 we plot
uncertainties of the ME and FE versus the real and imaginary parts
of $p$, both for the initial state
$|\Psi\rangle=(|00\rangle+|11\rangle)/\sqrt{2}$. Clearly, when $|p|$
is larger than a critical value $|p|_c$, the entanglement can always
be witnessed by TE, and the related entanglement region does not
explicitly depend on the particular choice of the reservoir. $|p|_c$
depends on the chosen initial state $\rho^{\psi}(0)$, and for
$|\Psi\rangle$ it is of about 0.9101, which gives the entanglement
region witnessed by it as $C^T\in[0.6858,1]$.

From Fig. 3 one can see that the ME and FE are determined, however,
by both ${\rm Re}(p)$ and ${\rm Im}(p)$, for there are $p_1$ and
$p_2$ (those with the same distance to the origin of the
coordinates) with $|p_1|=|p_2|$ but the uncertainties
$U^{(M,F)}(p_1)<1$ and $U^{(M,F)}(p_2)>1$ (here $U^{(M,F)}$
represents uncertainties estimated by ME and FE). This means that,
for this case, the entanglement regions witnessed by them are
determined by the explicit structures of the reservoir and the
coupling constant between the system and the reservoir. Of course,
if $p\in\mathbb{R}$ we still have the entanglement regions depending
only on $|p|$, and for the initial $|\Psi\rangle$ we obtained
numerically $|p|_c\simeq 0.9116$, $C^M\in[0.6905,1]$ for the  ME,
and $|p|_c\simeq0.9121$, $C^F\in[0.6921,1]$ for the FE.

For the initial $\rho^{\phi}(0)$, the density matrix
$\rho^{\phi}(t)$ is determined only by $|p|^2$, therefore all the
uncertainty estimates as well as the entanglement regions witnessed
by them are also determined by $|p|^2$, and thus are independent of
the explicit structure of the reservoir. In Fig. 2(b) we give an
exemplified plot of the $|p|$ dependence of different estimates for
the initial state $|\Phi\rangle=(|10\rangle+|01\rangle)/\sqrt{2}$,
for which we have $|p|_c\simeq 0.8962$, $C^T\in[0.8031,1]$ for the
TE, $|p|_c\simeq0.8982$, $C^M\in[0.8068,1]$ for the ME, and
$|p|_c\simeq0.8982$, $C^F\in[0.8486,1]$ for the FE.

In the following we give some explicit examples of the structured
reservoir to deepen our understanding of the above general
arguments.

\begin{center}
\textbf{A. The sub-Ohmic, Ohmic and super-Ohmic reservoirs}
\end{center}

We consider first the structured reservoirs with spectral densities
of the form \cite{Leggett}
\begin{equation}
 J(\omega)=\eta\omega^s\omega_c^{1-s} e^{-\omega/\omega_c},
\end{equation}
with $\eta$ and $\omega_c$ being the dimensionless coupling constant
and the cutoff frequency, which are related to the reservoir
correlation time $\tau_B$ and the relaxation time $\tau_R$ (over
which the state of the system changes in the Markovian limit of flat
spectrum) by $\tau_B\approx\omega_c^{-1}$ and
$\tau_R\approx\eta^{-1}$. Depending on the value of $s$, the
reservoir is classified as sub-Ohmic if $0<s<1$, Ohmic if $s=1$, and
super-Ohmic if $s>1$.

For this kind of reservoir spectral densities, $p(t)$ is determined
by \cite{Tongqj}
\begin{equation}
 \dot{p}(t)+i\omega_{0}p(t)+\int_0^t p(t_1)f(t-t_1)dt_1=0,
\end{equation}
where the kernel function $f(t-t_1)=\int d\omega
J(\omega)e^{-i\omega(t-t_1)}$ in the continuum limit the spectral
density.

In this work we take $s=1/2$, 1 and 3 as three examples of the
sub-Ohmic, Ohmic, and super-Ohmic spectral densities. The kernel
function can be integrated as $f(x)=\eta
s!\omega_c^2/(1+i\omega_cx)^{s+1}$ ($x=t-t_1$) for $s\in\mathbb{Z}$,
and $f(x)=\eta\omega_c^2\sqrt{\pi}e^{-i\varpi}/[2(1+\omega_c^2
x^2)^{3/4}]$ for $s=1/2$, where $s!$ denotes the factorial of $s$,
and $\varpi=\frac{3}{2}\tan^{-1}(\omega_c x)$. Then $p(t)$ can be
solved numerically and the two-qubit density matrix can be derived
by the procedure of \cite{Bellomoprl}.

\begin{figure}
\centering
\resizebox{0.45\textwidth}{!}{%
\includegraphics{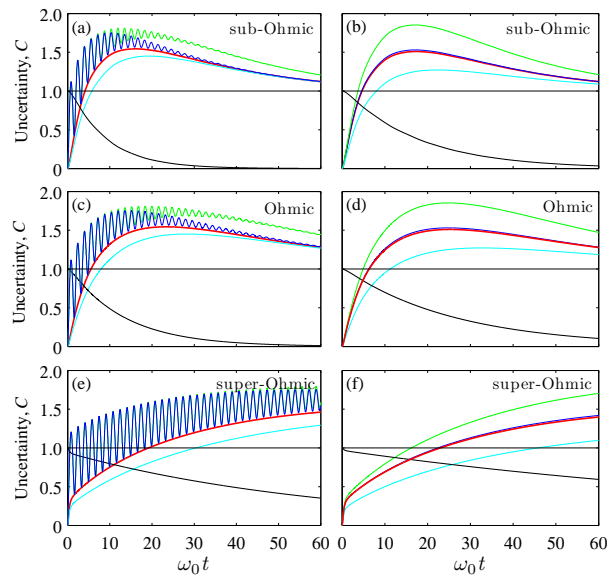}}
\caption{(Color online) Entanglement witness in the sub-Ohmic, Ohmic
and super-Ohmic reservoirs for the initial $|\Psi\rangle$ (left) and
$|\Phi\rangle$ (right), with $\eta=0.01$, $\omega_c=2\omega_0$. The
green, blue, red, cyan and black lines from top to bottom represent
FE, ME, TE, BB, and the concurrence, respectively.} \label{fig:4}
\end{figure}

In Fig. 4 we presented theoretical predictions of entanglement
witness for the initial $|\Psi\rangle$ and $|\Phi\rangle$ in
sub-Ohmic, Ohmic and super-Ohmic reservoirs, with $\eta=0.01$ and
$\omega_c=2\omega_0$. One can see that while the concurrence decays
monotonically and disappears in the infinite time limit, the
different uncertainty estimates become larger than 1 after finite
timescales. Thus there are critical $\omega_0 t_c$ after which the
entanglement cannot be witnessed by them. For the TE, although the
time interval $\omega_0 t_c$ is different for different spectral
densities of the reservoirs and different system-reservoir coupling
constants, the numerical results confirm that the entanglement
regions witnessed by it agree with those predicted in the general
arguments (i.e., $C^T\in[0.6858,1]$ for the initial $|\Psi\rangle$,
and $C^T\in[0.8031,1]$ for the initial $|\Phi\rangle$).

For the ME and FE, the regions of entanglement being witnessed will
shrink, particularly, for the initial $|\Psi\rangle$, both the ME
and FE oscillate around 1 with increasing $\omega_0 t$ in the
short-time region (during which they are nearly overlapped), and the
entanglement regions witnessed by them are discontinuous because
here $p(t)\in\mathbb{C}$. For example, for parameters of Fig. 4(a)
the discrete entanglement regions are $C^M\approx
C^F\in[0.7501,0.7897],~[0.8748,0.9491]$ and $[0.9794,1]$; that is,
some states with a small amount of entanglement can be witnessed,
while some others with a relatively large amount of entanglement
cannot be witnessed. This reveals a counterintuitive fact; that is,
not the more the state is entangled, the easier it can be witnessed
for some schemes. For the initial $|\Phi\rangle$, the related
entanglement regions are in accord with those predicted in the above
general arguments, which are independent of the explicit time
dependence of $p(t)$.

We also examined effects of the cutoff frequency $\omega_c$ on
entanglement witness (for concise presentation in the paper, we did
not plot them here), and found that for the sub-Ohmic and Ohmic
spectral densities, $\omega_0 t_c$ decreases with increasing
$\omega_c$ and their dependence on $\omega_c$ are weak ($\omega_0
t_c$ for $|\Phi\rangle$ is only a slightly larger than that for
$|\Psi\rangle$). But for the super-Ohmic spectral density, $\omega_0
t_c$ is strongly dependent on $\omega_c$; for example, for the
parameters of Figs. 4(e) and 4(f), they increase dramatically from
8.125 and 9.55 for $\omega_c=\omega_0$ to 165.525 and 196.85 for
$\omega_c=6\omega_0$.

\begin{center}
\textbf{B. The Lorentzian reservoir}
\end{center}

As the second example, we consider the structured reservoir with the
Lorentzian spectral density \cite{Breuer}
\begin{equation}
 J(\omega)=\frac{1}{2\pi}\frac{\gamma_0\lambda^2}{(\omega-\omega_c)^2+\lambda^2},
\end{equation}
where $\lambda$ denotes the spectral width of the reservoir and is
related to the reservoir correlation time via
$\tau_B\approx\lambda^{-1}$, and $\gamma_0$ is related to the
relaxation time $\tau_R$ via $\tau_R\approx\gamma_0^{-1}$.
$\lambda>2\gamma_0$ ($\lambda<2\gamma_0$) corresponds to the
Markovian (non-Markovian) regime, and $\omega_c=\omega_0-\delta$ is
the central frequency of the reservoir detuned from the transition
frequency $\omega_0$ by an amount $\delta$.

The function $p(t)$ can be derived analytically as \cite{Hufan}
\begin{equation}
 p(t)=e^{-\frac{1}{2}(\lambda-i\delta)t}\left[\cosh\frac{dt}{2}
 +\frac{\lambda-i\delta}{d}\sinh\frac{dt}{2}\right],
\end{equation}
with $d=[(\lambda-i\delta)^2-2\gamma_0\lambda]^{1/2}$. Thus the
two-qubit density matrix can also be derived analytically
\cite{Bellomoprl}.

\begin{figure}
\centering
\resizebox{0.4\textwidth}{!}{%
\includegraphics{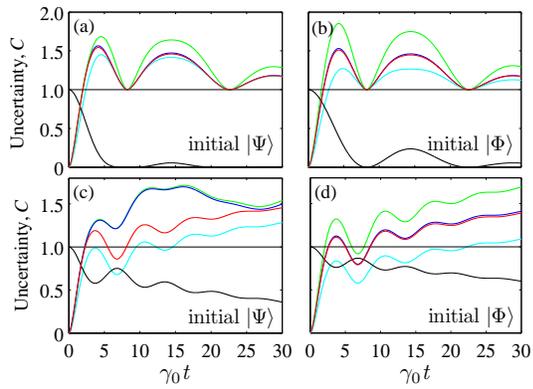}}
\caption{(Color online) Entanglement witness in Lorentzian reservoir
with the initial states $|\Psi\rangle$ and $|\Phi\rangle$, where
$\lambda=0.1\gamma_0$, $\delta=0$ for panels (a) and (b),
$\delta=0.8\gamma_0$ for panels (c) and (d). The green, blue, red,
cyan and black lines from top to bottom represent FE, ME, TE, BB,
and the concurrence, respectively.} \label{fig:5}
\end{figure}

\begin{figure}
\centering
\resizebox{0.4\textwidth}{!}{%
\includegraphics{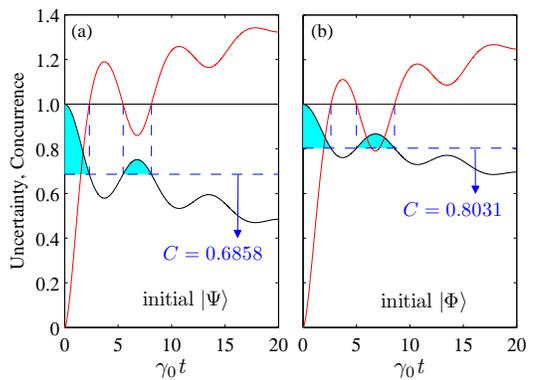}}
\caption{(Color online) Comparison of the entanglement (black)
witnessed by TE (red) at different time regions, where the two
panels are plotted with completely the same parameters as those in
Figs. 5(c) and 5(d), respectively.} \label{fig:6}
\end{figure}

To explore efficiency of entanglement witness for this kind of
system-reservoir coupling, we plot in Fig. 5 theoretical predictions
of uncertainties in the non-Markovian regime ($\lambda=0.1\gamma_0$)
with the initial $|\Psi\rangle$ and $|\Phi\rangle$. Clearly, they
oscillate with increasing $\gamma_0 t$, and the time regions during
which the entanglement can be witnessed depend on $\delta$ and
$\lambda$. But in the absence of detuning (i.e., $\delta=0$), the
numerical results show that the entanglement regions witnessed by
different estimates are independent of $\lambda$, and are consistent
with those predicted in the above general arguments, which can be
understood from the fact that when $\delta=0$ we always have
$p(t)\in\mathbb{R}$.

Introducing detuning will decrease the decay rate of entanglement,
and as can be seen from Figs. 5(c) and 5(d), the time regions during
which the entanglement can be witnessed become discontinuous, even
for the TE. But the entanglement regions witnessed by TE remains the
same (see the green shaded regions in Fig. 6) as that for
$\delta=0$. Furthermore, for the initial $|\Psi\rangle$, the
entanglement regions witnessed by ME and FE vary with the variation
of $\lambda$ (their dependence on $\lambda$ may be very weak for
certain parameters, e.g., $\delta=0.8\gamma_0$ and
$\lambda>0.5\gamma_0$), which is caused by $p(t)\in\mathbb{C}$ when
$\delta\neq 0$. For the initial $|\Phi\rangle$, they remain the same
as those for $\delta=0$, and do not depend on the parameter
$\lambda$.

\section{Summary}
In summary, we have studied relations between the
quantum-memory-assisted entropic uncertainty principle,
teleportation and entanglement witness. We proved geometrically that
any two-qubit state with negative conditional von Neumann entropy,
which thus lowers down the upper bound of the entropic uncertainty
relation, is useful for teleportation (i.e., $F_{\rm av}>2/3$). We
have also examined efficiency of this new entropic uncertainty
principle on witnessing entanglement in a general class of bosonic
structured reservoirs, and found that the entanglement regions
witnessed by TE for the initial EWL state $\rho^{\psi}(0)$ or that
witnessed by all the three estimates TE, ME and FE for the initial
EWL state $\rho^{\phi}(0)$ are determined only by a function $p$,
which has no relation with the explicit form of its time dependence.
These general arguments are corroborated by explicit examples of
structured reservoirs with the sub-Ohmic, Ohmic, super-Ohmic and
Lorentzian spectral densities. As a by-product, we also found that
it is not a general result that the more the state is entangled, the
easier it can be witnessed for certain chosen schemes.

As the quantum-memory-assisted entropic uncertainty principle has
been experimentally realized \cite{Prevedel,Licf}, and it is
possible to simulate and control the Markovian and non-Markovian
environments \cite{Almeida,Xujs,Liubh}, we expect the results
demonstrated in this work may be certified in future experiments
with currently available technologies; for example, by using
two-level atoms confined in optical microcavities \cite{Vahala} or
simulated reservoirs \cite{Buluta}.

\begin{center}
\textbf{ACKNOWLEDGMENTS}
\end{center}

This work was supported by NSFC (11205121, 10974247, 11175248),
``973'' program (2010CB922904), NSF of Shaanxi Province
(2010JM1011), and the Scientific Research Program of Education
Department of Shaanxi Provincial Government (12JK0986).

\newcommand{\PRL}{Phys. Rev. Lett. }
\newcommand{\PRA}{Phys. Rev. A }
\newcommand{\JPA}{J. Phys. A }
\newcommand{\JPB}{J. Phys. B }
\newcommand{\PLA}{Phys. Lett. A }
%

%

\end{document}